\documentclass[11pt]{article}
\usepackage{jheppubmod}
\usepackage{hyperref}
\pdfoutput=1
\usepackage{psfrag}
\usepackage{array}
\usepackage{amssymb}
\usepackage{amsmath}
\usepackage{amsthm}
\usepackage{float}
\usepackage{graphicx}
\usepackage{caption}
\usepackage{color}

\usepackage{epsfig}
\def\pb[#1,#2]{\{#1, #2\}}
\def\deb[#1,#2]{[#1,#2]_{\text{D.B.}}}

\def\Or[#1]{{\text{O}}\left({#1}\right)}
\def\dotl[#1,#2]{\left\langle #1,\, #2 \right\rangle}
\def\dotlb[#1,#2]{\left\langle #1,\, #2 \right\rangle}
\def\dotlm[#1,#2]{\left[ #1,\, #2 \right]}
\def\dotp[#1,#2]{(\vect{#1} \cdot\vect{#2})}
\def\aff[#1,#2]{\hat{#1}(#2)}

\def\n4sym{{\cal N}=4 SYM}
\def\>{\rangle}
\def\<{\langle}
\def\weight[#1,#2,#3]{\{(#1),#2,#3\}}
\def\ads[#1]{$\text{AdS}_{#1}$}

\newcommand{\be}{\begin{equation}}
\newcommand{\ee}{\end{equation}}
\newcommand{\ba}{\begin{align}}
\newcommand{\ea}{\end{align}}
\newcommand{\bs}{\begin{split}}
\def\sess\end{split}
\newcommand{\vect}[1]{{\boldsymbol{#1}}}

\title{A note on AdS cosmology and gauge theory correlator }
\author[a]{Souvik Banerjee} 
\affiliation[a]{Van Swinderen Institute for Particle Physics and Gravity, University of Groningen,
Nijenborgh 4, 9747 AG, The Netherlands.}
\author[b]{Samrat Bhowmick}
\affiliation[b]{Department of Physics, Indian Institute of Technology Madras, Chennai, India - 600 036.}
\author[c]{Soumyabrata Chatterjee}
\author[c]{Sudipta Mukherji}
\affiliation[c]{Institute of Physics, Sachivalaya Marg, Bhubaneswar, India - 751 005.}
\emailAdd{souvik.banerjee@rug.nl}
\emailAdd{samrat@physics.iitm.ac.in}
\emailAdd{soumyac@iopb.res.in}
\emailAdd{mukherji@iopb.res.in}

\date{January, 2015}

\abstract{Using AdS/CFT prescription, we compute two point Yang-Mills correlator on a constant time slice 
for the Kasner background. Pushing the surface close to the initial singularity, we find, in some cases, 
the correlator does not develop pole. We further numerically calculate similar correlator where the bulk is a Kasner 
AdS soliton. We find that the qualitative behaviour of the correlator remains unchanged.}
\begin{document}
\maketitle
\section{Introduction}
AdS/CFT correspondence relates string theory in AdS space to a supersymmetric $SU(N)$ Yang-Mills theory living on the 
boundary of this AdS. The isometry group of the AdS orders the boundary field theory to be conformal. The correspondence
can be summarized through the relation :

\begin{equation}
\frac{L_{AdS}^4}{l_s^4} = 4 \pi g_{YM}^2 N \equiv 4 \pi \lambda,
\end{equation}
where $l_s$ is the string length and $L_{AdS}$, the curvature length scale of the AdS. $g_{YM}$ and $N$ denote the coupling 
constant and the rank of the gauge group of the boundary gauge theory respectively. $\lambda = g_{YM}^2 N$ is the 
t'Hooft coupling. In the limit, $N \rightarrow \infty$,  $g_{YM} \rightarrow 0$ and $\lambda = \text{finite}$ the string 
theory is classical. When the t'Hooft coupling is large ($\lambda >> 1$), the classical string theory reduces to
its low energy supergravity limit. Furthermore, a large $\lambda$ makes the field theory to be strongly coupled while
the bulk gravity theory remains weakly coupled and hence AdS/ CFT correspondence reduces to a weak-strong duality
between a classical gravity theory and a quantum field theory. However, due to strong gravitational fluctuations, this 
simple limit of the correspondence ceases to exist when the boundary hits, for example, cosmological singularities.
Attempts were made to use AdS/CFT to construct time-dependent bulk geometries dual to a boundary field theory well
defined even near the cosmological singularity. This is something expected from the fact that AdS/ CFT correspondence can be 
thought of as an example of a more general open string- closed string duality in string theory. This means, even if the 
classical supergravity limit of string theory does not exist near singularity, it, in principle, does
not prevent one from having a well-defined dual field theory right there. However, the successes in constructing 
such holographic models have been partial \cite{Hertog:2005hu} - \cite{Awad:2009bh}.

In this regard, a recent work \cite{Engelhardt:2014mea} has provided us with a holographic approach
to examining physics near a class of cosmological singularities\footnote{Earlier, in \cite{Koyama:2001rf},
boundary correlators were calculated by embedding time dependent backgrounds in AdS.} The 
boundary here is given by the Kasner geometry
which can appropriately be extended to the bulk. This bulk metric solves five dimensional Einstein equation 
in the presence of a negative cosmological constant.  
The work  then uses space-like correlators of CFT  operators with high enough scaling dimension living on the boundary as a probe.  
These are ``good'' probes because the correlators can be well-approximated by exponentiated length of
bulk geodesics provided the geodesics go close enough to the cosmological singularity. It is then
shown that in certain circumstances the correlator develops a pole in their IR behaviour whenever the geodesic approaches the 
$t =0$ singularity of the Kasner geometry.

Inspired by this work, in this short note, we revisit the calculation of the correlators. 
We, however, take a different route of computation. Instead of calculating it on a {\it 
fixed}-time surface, we exploit the underlying scaling symmetry of the Kasner AdS space-time to evaluate the equal-time correlator. The whole process can equivalently be thought of as evaluating the same on an arbitrary space-like surface which finally we  push all the way to cosmological singularity and extract the nature of the correlator therein.
We find that in the cases where the two points in the boundary are separated in the spatial direction associated with a positive Kasner exponent \footnote{Given the Kasner conditions, namely,$\sum_{i} p_i =1$ and $\sum_{i}p_i^2=1$, $p_i$'s being the Kasner exponents, it is guaranteed that we always have {\it{at least}} one positive Kasner exponent.}, no pole appears in the correlator even near
the cosmological singularity.
Throughout our analysis, the said asymmetric scaling symmetry of the background metric remains manifest. This in turn
enables us to  study some general features of the correlator as well.

While the above  bulk Kasner-AdS geometry has a Poincare horizon at the 
core, it is possible to cap it off at finite radial distance by considering a Kasner-AdS soliton instead. We 
numerically compute the same correlator for this new background and find that the qualitative behaviour remains
unchanged.

The plan of the paper is as follows.
In the next section, we give a brief account of the cosmological metrics with anisotropic scaling symmetry
including the Kasner metric. In section 3, we compute the correlator and study it's properties in a specific
cases. In the  appendix A,  we discuss how the relevant time-dependent geometries can be obtained as near horizon 
limits of 
brane-solutions in supergravity. Calculational details are presented in the appendix B. 

\section{Scaling Solutions}
Several time-dependent solutions of five-dimensional Einstein equations in the presence of negative cosmological constant
are known which have anisotropic scaling symmetry of the form
\begin{equation}
z \rightarrow \lambda z, ~~t \rightarrow \lambda t, ~~ x_i \rightarrow \lambda^{(1-p_i)} x_i ~(i = 1,2 ...,n),
\label{suscal}
\end{equation}
where $p_i$'s and $\lambda$ are some constants. Here are some of the examples:

%\subsubsection*{Examples}
\begin{itemize}
 \item The most celebrated example of this comes in the form 
 \begin{equation}
ds^2 = \frac{1}{z^2} \Big[ - dt^2 + dz^2 + \sum_{i=1}^{n} (a_i)^2 dx_i^2\Big],
\label{suone}
\end{equation}
where $a_i = t^{p_i}$. $p_i$ are known as  Kasner exponents and satisfies the Kasner conditions, namely, $\sum_{i=1}^{n} p_i = 1$ and  
$\sum_{i=1}^{n} p_i^2 = 1$
in order that (\ref{suone}) is a $AdS_{n+2}$  space-time. For $n=3$ and $n=5$ these solutions can also be realized from the perspective 
of $10$ and $11$ dimensional 
supergravity theories as near horizon $D3$  and $M5$ brane solutions with Kasner like world volume. Such solutions and their 
cosmological implications have been
studied in detail in \cite{Banerjee:2013jn}.

\item One of the Kasner conditions can however be relaxed at the expense of introducing matter fields. For instance, there exist $AdS$ 
solutions with the same metric 
(\ref{suone}) with $\sum p_i = 1$.  For that one needs to introduce a dilatonic 
scalar field, $\Phi = \lambda \, Log 
\, t$, 
where $\lambda$ is defined through 
the modified Kasner condition, $\sum_{i=1}^{n} p_i^2 = 1 - \frac{\lambda^2}{2}$. 
Such solutions were studied in \cite{Awad:2007fj} in 
the context of probing cosmological 
singularity through gauge theory duals. These solutions have the scaling symmetry as long as dilaton is shifted
by a appropriate constant.
 
\item Finally we shall give example of another class of 
solutions which has slightly different form than (\ref{suone}) but still obeys the scaling 
relations, (\ref{suscal}). The metric of the solution takes the form  
\begin{equation}
\label{kasol}
 ds^2 = \frac{1}{z^2}\left[-d{t}^2 + \frac{1}{f(z)} dz^2 + \sum_{i=1}^{n-1} t^{2 p_i} dx_i^2 + f(z) dx_n^2\right]
\end{equation}
Here $p_i$'s satisfy Kasner conditions with $p_n = 0$ and $f(z) = 1-\frac{z^{n+1}}{z_0^{n+1}}$. The coordinate, $x_n$ is an angular coordinate 
here.
These solutions are known as Kasner $AdS_{n+2}$ solitons \cite{Horowitz:1998ha, Engelhardt:2013jda} and 
can be realized in supergravity (for $n=5, 7$). We put the supergravity origin of these solutions
in the appendix A as this is beyond the main point of discussion of this work.
\end{itemize}

\section{Gauge Theory on Time-Dependent Boundary : Consequences}
We now turn to the dual gauge theory. As mentioned in the introduction, one of the prime motivations of 
studying time-dependent $AdS$ geometries have been to understand the physics near cosmological singularity, $t \rightarrow 0$, a sector,
otherwise intractable from a direct study.

In order to proceed, one generally computes the space-like two point 
correlator, $\langle \psi| {\cal O}(x, t) {\cal O}(x', t) |\psi\rangle$ on a state $|\psi\rangle$ of the strongly coupled 
Yang-Mills theory residing on the boundary of some of the bulk geometries we discussed in the previous section.
When the boundary CFT has a well defined large $N$ limit,
the correlator, in the leading order, can be well approximated by
\begin{equation}
\langle \psi| {\cal O}(x) {\cal O}(x') |\psi\rangle = e^{-m {{\cal L}_{reg}} (x,x')},
\label{sureg}
\end{equation}
provided the operator, $\cal O$ has a high scale dimension, $\Delta = \frac{d}{2} + \sqrt{\frac{d^2}{4} + m^2}$, $d$ being the 
boundary dimension.
Here ${\cal L}_{reg}$ is the regularized length of the geodesic whose end points are fixed at boundary points, $x$ and $x^\prime$. With this definition in mind, in the rest of this section we shall compute two examples. First we shall revisit the example of Kasner $AdS$ space-time and then we shall move on to solitonic Kasner - $AdS$ . We will show how some generic features of the result emerge as a consequence of the underlying scaling symmetry of those solutions. 
 For Kasner-Ads soliton, owing to the complication due to
the background geometry, we compute the correlator
numerically and compare the result with that of Kasner-AdS.

\subsection{Example 1 : $AdS$ - Kasner}
We would like to compute $\langle {\cal O}(x_1', t_0) {\cal O}({x_1}'', t_0) \rangle$ where the bulk geometry 
is given by (\ref{suone})
with $a_i = t^{p_i}$. This is a 
correlator along $x_1$ direction with two boundary points at $x_1', {x_1}''$ computed
at a fixed time $t = t_0$. Corresponding space-like geodesic must then have two fixed end
points $x'_1, {x''}_1$ at the boundary $z =0$ at time $t = t_0$. For this particular
calculation, therefore, the other boundary directions  $x_i, i \ne 1$ are irrelevant. 
For the moment, we work with a general scale factor $a_1(t)$ along $x_1$. Later, we will
use the explicit form $a_1 = t^{2p_1}$ for $p_1 >0$.

Calling $x_1$ as $x$ and $a_1$ as $a$ for notational simplicity, the geodesic equations for (\ref{suone})
are given by
\begin{eqnarray}
&&x^{\prime\prime} + 2 \frac{a^\prime}{a} x^\prime - a a^\prime {x^\prime}^3 = 0,\nonumber\\
&&z z^{\prime\prime} + {z^\prime}^2 + {x^\prime}^2 a^2 - a a^\prime z z^\prime {x^\prime}^2 -1 = 0.
\label{suthree}   
\end{eqnarray}
Here, we have taken time as a parameter and derivatives are with respect to time.

General solutions of these equations can be written as
\begin{equation}
x (t) = \pm \int \frac{a(t^*) dt}{a(t) {\sqrt{a^2(t^*) -a^2(t)}}}
\label{sufour}
\end{equation}
and
\begin{equation}
z = + {\sqrt{ -2 \int dt  \big[\frac{a(t)}{\sqrt{a^2(t^*) - a^2(t)}} \Big(\int^{t} dt' \frac{a(t')}{\sqrt{a^2(t^{\prime *}) - 
a^2(t')}}\Big)\Big]
}}.
\label{sufive}
\end{equation}
To write (\ref{sufour}) in this form, we have used the fact that there is a turning point of the
geodesic in the bulk and at that point $dx/dt$ diverges. For the solution above, we have taken
the point to be $t = t^*$.

Given a functional form for $a(t)$, one would then try to integrate the left hand sides
of the above equations. In this process, three integration constants would appear. However,
all of these can be fixed by boundary conditions. The constant appearing from (\ref{sufour})
can be set to zero by using $x \rightarrow x + {\rm constant}$ symmetry of the metric.
Other two constants arise from the two integrations in (\ref{sufive}).
Both of them can be fixed - (1) by demanding $dz/dx =0$ at the turning point of the geodesic in $z-x$
plane and (2) by requiring $z =0$ for $t = t_0$.

We now argue that if the metric has the scaling symmetry (\ref{suscal}), then the constant 
$a(t^*)$ can be scaled away. Taking $a(t) = t^p$ and defining new coordinates
$\bar z, \bar t$ and $\bar x$ as
\begin{equation}
z = t^* \bar z,~~t = t^* \bar t, ~~x = {t^*}^{1-p} \bar x
\label{susix}
\end{equation}
we can re-write (\ref{sufour}) and (\ref{sufive}) as
\begin{equation}
\bar x (\bar t) = \pm \int \Big[\frac{d\bar t}{ {\bar t}^p  {\sqrt{1 - \bar t^{2p}}}}\Big],
\label{suseven}
\end{equation}
and
\begin{equation}
\bar z(\bar t) = + {\sqrt{ -2 \int  d\bar t \Big[ \frac{\bar t^p} {\sqrt{1 - \bar t^{2p}}}
\Big( \int^{\bar t}  d\bar t' \frac{{\bar t}'^p} {\sqrt{1 - {\bar t}'^{2p}}}\Big)\Big] }}.
\label{sueight}
\end{equation}

With this, (\ref{suseven}) and (\ref{sueight}) can be easily integrated. This gives, for {\it generic} $p$
\begin{eqnarray}
\bar x(\bar t) = \frac{{\bar t}^{1-p}}{1 - p} ~{}_2F_1 \Big(\frac{1}{2}, \frac{1-p}{2p},
\frac{1+p}{2p}, {\bar t}^{2p}\Big) - \frac{\sqrt{\pi} \Gamma(\frac{1-p}{2p})}{(1-2p) \Gamma(\frac{1-2p}{2p})},
\label{sunine}
\end{eqnarray}
and 
\begin{eqnarray}
\bar z(\bar t) = && \Big[{\bar t}^2\Big[ 1 - {}_3F_2\Big( \{1, \frac{1}{2p}, \frac{1}{p}\}, \{\frac{1}{2} + 
\frac{1}{2p}, 1 + \frac{1}{p}\}, {\bar t}^{2p}\Big)\Big] \nonumber\\
&&+\frac{4{\sqrt \pi} p \Gamma(\frac{1+p}{2p})
{\bar t}^{1-p} } { (1-2p)\Gamma(\frac{1-2p}{2p})}
\Big[ {}_2F_1\Big(\frac{1}{2}, \frac{1-p}{2p}, \frac{1+p}{2p}, {\bar t}^{2p}\Big)
- {\sqrt {1 - {\bar t}^{2p}}}\Big]
+ c\Big]^{\frac{1}{2}}.
\label{suten}
\end{eqnarray}   
Here, $c$ is a constant which can fixed using $\bar z = 0$ for $\bar t = \bar t_0$.
${}_2F_1$ and ${}_3F_2$ are the hypergeometric function and the generalized hypergeometric 
function respectively. For some specific values
of $p$, the solutions however simplify. 
In appendix B, we provide a way to solve (\ref{suthree}) and get to 
these results.

Having reached this far, we proceed to find the geodesic length.
For the correlator, $\langle {\cal O}(x, t_0) {\cal O}(-x, t_0)\rangle$, we first need
to calculate the integral (\ref{sureg})
\begin{equation}
{\cal L} = \int \frac{2 dt}{z} \Big[{\sqrt{ -1 + \Big(\frac{dz}{dt}\Big)^2 + t^{2p} \Big(\frac{dx}{dt}\Big)^2}}\Big],
\end{equation}
with appropriate limits. Now, as for the lower limit, the turning point of the geodesic is at $t = t^*$.
In terms of scaled time, it is at $\bar t = 1$. For the upper limit, we note that the correlator is being 
calculated at a constant $t = t_0$ slice. This, in terms of scaled variable, is $ \bar t = \bar t_0$. 
We further  need to UV-regulate the integral by introducing a cut-off, $\bar \delta$. The geodesic length is therefore
\begin{equation}
{\cal L} = \int_{\bar t =1}^{\bar t = \bar t_0 - \bar \delta}  \frac{2 d\bar t}{\bar z} \Big[{\sqrt{ -1 + \Big(\frac{d\bar 
z}{d\bar t}\Big)^2 + {\bar t}^{2p} \Big(\frac{d\bar x}{d\bar t}\Big)^2}}\Big].
\end{equation}

In general, $\cal L$ is infinite. In order to render it finite, we need to subtract, from $\cal L$, the 
equivalent AdS part. This removes
the $\bar \delta \rightarrow 0$ singularity in the geodesic length. Consequently, the regulated $\cal L$ will
only depend on $\bar t_0$.

Above observation, in turn, means that the gauge theory correlator has the form
\begin{equation}
\langle  {\cal O}(x, t_0) {\cal O}(- x, t_0) \rangle = e^{-m {\cal{L}}_{reg}} = f(\bar t_0),
\label{sueleven}
\end{equation}
for some function $f$. For general $p$, we are unable to evaluate this function analytically. Nevertheless, numerically 
it can be calculated. We provide our results later. However,
for some values of $p$, expressions simplify and analytic computations can be done. As an illustrative example,
we do it for $p = 1/3$. Results are given below.  
\begin{eqnarray}
&& \bar z(\bar t) = {\sqrt{ 3 ({\bar t}^{~\frac{4}{3}} - {\bar t}_0^{~\frac{4}{3}}) + ({\bar t}^2 - {\bar t}_0^2)}}\nonumber\\
&&\bar x(\bar t)=  \pm 3 {\sqrt{1 - {\bar t}^{~\frac{2}{3}}}}.
\end{eqnarray}
The geodesic length turn out to be
\begin{eqnarray}
{\cal L} &&= \int_{\bar t =1}^{\bar t_0 - \bar \delta} d\bar{t}
\frac{  2 \bar t^{\frac{1}{3}} {\sqrt{1 - {\bar t_0}^{~\frac{2}{3}}}} (2 + {\bar t_0}^{~\frac{2}{3}}) }
{ {\sqrt{1 - {\bar t}^{~\frac{2}{3}}}} ( 3 {\bar t}^{~\frac{4}{3}} + {\bar t}^{~2} - 3 {\bar t_0}^{~\frac{4}{3}}
- {\bar t_0}^{~2})}\nonumber\\
&& = 2 ~{\rm tanh}^{-1}~\left[ \frac{ {\sqrt{ 1 - ({\bar t_0} - {\bar\delta})^{\frac{2}{3}} }}
\,\left(2 + (\bar t_0 -\bar \delta)^{\frac{2}{3}}\right) }
{ {\sqrt{1 - {\bar t}_0^{\frac{2}{3}} }} \left(2 + {\bar t}_0^{~\frac{2}{3}}\right) }\right].
\end{eqnarray}
Finally, to  obtain the regularized length, we need to subtract appropriate AdS contribution. Therefore,
\begin{eqnarray}
{\cal L}_{reg} &&= {\cal L} - 2 ~{\rm log}~\left[\frac{ {\bar t_0}^{~\frac{1}{3}}} { \bar z(\bar t_0 - \bar \delta)}\right]\nonumber\\
&&= {\rm log}~\left[ \frac{ 4 ( 4 -{\bar t}_0^2 - 3 {\bar t}_0^{~\frac{4}{3}})}
{{\bar t}_0^{~\frac{2}{3}}}\right].
\end{eqnarray}
Therefore, we find that $f(\bar t_0)$ goes to zero as we take $\bar t_0 \rightarrow 0$. 

% The smooth behaviour of the two point correlator at this cosmological singularity for
% $p >0$ is perhaps not so surprising. Our result shows, as we go towards singularity field
% theory becomes weakly coupled, and that is expected because near cosmological singularity
% gravity becomes strong.

As we mentioned previously, for arbitrary $p$, it is not possible to evaluate the correlator analytically. However, 
it is straightforward to carry out a numerical computation. The result is shown in figure 1.
Indeed, we find the correlators do not pick up singularities as we take ${\bar t}_0 \rightarrow 0$.
\begin{figure}[H]
\label{fig1}
 \centering
 \includegraphics[width=.60\textwidth]{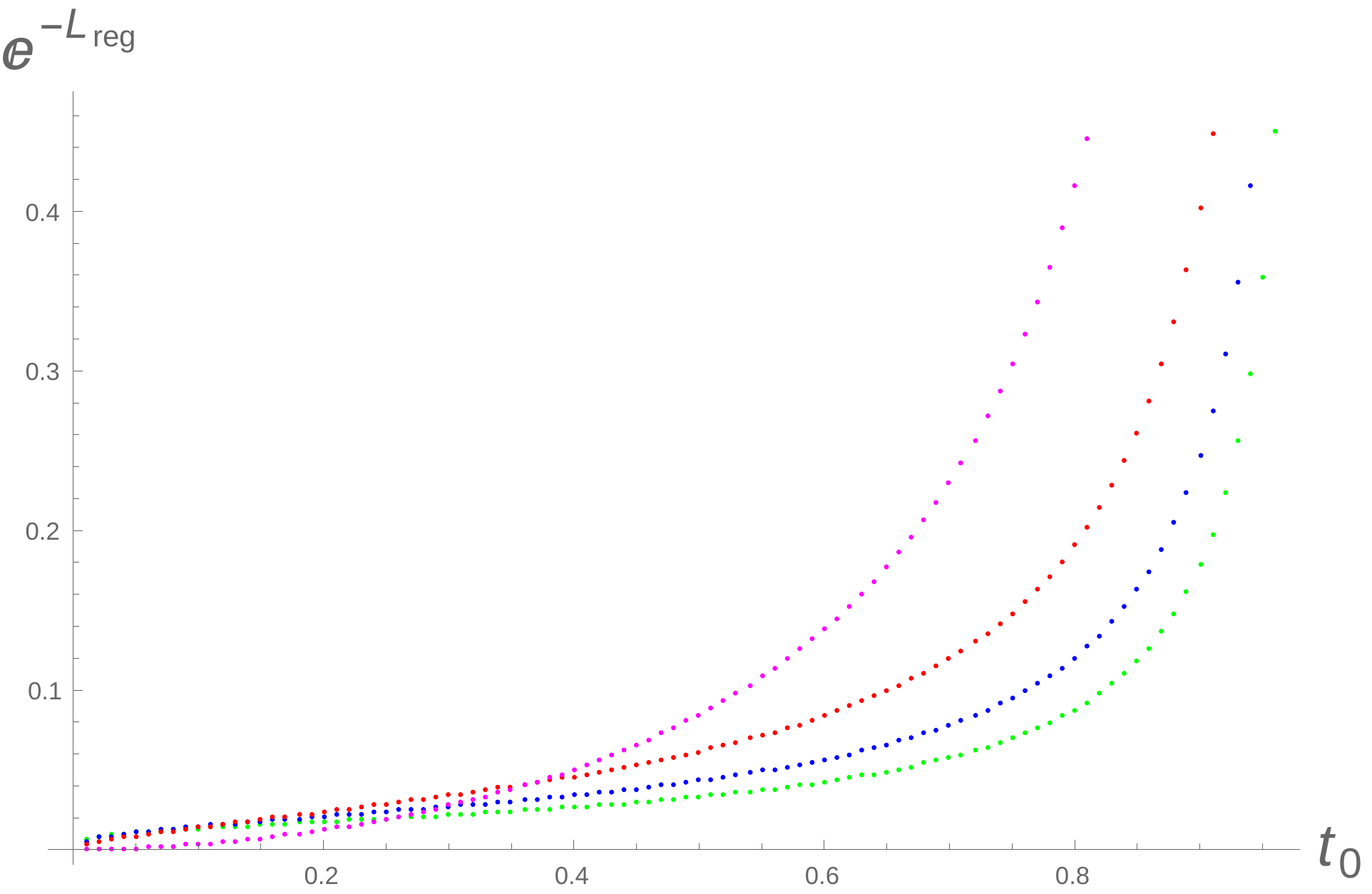}
 \caption{Plot of $e^{-L_{reg}}$ vs $t_0$ for different values of $p$.  Magenta, blue and green 
are for $p = 9/10, 1/5$ and $1/7$ respectively. For $p =1/3$, the numerical and analytical results
coincide. The behaviour is shown above in red.}
\end{figure}

Having computed the correlator for $p >0$, we proceed to make some general remarks about the correlator. 
First, let us notice that we can re-write (\ref{sueleven}) as
\begin{equation}
\langle  {\cal O}(t^{* (1-p)} \bar x, t^*{\bar t}_0) {\cal O}(-t^{* (1-p)} {\bar x}, t^*{\bar t}_0) \rangle = f(\bar t_0).
\label{sutwelvea}
\end{equation}
However, since $t^*$ is a free parameter, we are free to choose it. Let us take 
$t^* = {\bar t_0}^{-1}$. Then the correlator takes the form
\begin{equation}
\langle  {\cal O}({\bar t}_0^{(p-1)} \bar x, 1)  {\cal O}(-{\bar t}_0^{(p-1)} \bar x, 1) \rangle = f(\bar t_0).
\end{equation}
Calling ${\cal O}({\bar t}_0^{(p-1)} \bar x, 1) = {\cal {\tilde O}} ({\bar t}_0^{(p-1)} \bar x)$ and so on
we get,
\begin{equation}
\langle {\cal {\tilde O}} ({\bar t}_0^{(p-1)} \bar x) {\cal {\tilde O}} (-{\bar t}_0^{(p-1)} \bar x)\rangle = f(\bar t_0).
\label{sutwelve}
\end{equation}
Dependence on the arguments of correlator in this fashion is indeed expected in a scale invariant
theory. Note that for $p >0$, as we push the space-like surface close to $\bar t_0 =0$, 
the separation between the two points in the correlator increases. So we capture the large separation
behaviour of the correlator. 

The second scaling solution of the previous section has similar metric but there is a 
non-trivial dilaton. Though this scalar goes to zero at $t =0$, it diverges at a later time
-- leading to the divergence in Yang-Mills coupling. This however is not a concern for the 
third scaling solution. This is the Kasner-AdS soliton and we now compute the 
gauge theory correlator at the boundary of this background.

\subsection{Example 2 : Kasner soliton in AdS}
The $AdS_7$ Kasner soliton is given by
\begin{eqnarray}
\label{M5nhz1}
 ds^2 &=& \frac{1}{z^2}\left[-d\bar{t}^2 + \bar{t}^{2 \alpha_1} d\bar{x}_1^2 
+ \bar{t}^{2 \alpha_2} d\bar{x}_2^2 + \bar{t}^{2 \alpha_3} d\bar{x}_3^2 
+ \bar{t}^{2 \alpha_4} d\bar{x}_4^2 +\left(1-\frac{z^6}{z_0^6}\right)
d\bar{\theta}^2 + \left(1-\frac{z^6}{z_0^6}\right)^{-1} dz^2\right]\nonumber\\
\end{eqnarray}
This is precisely the $n=5$ case  of (\ref{kasol}).
The details of derivation of this form is given in (\ref{M5nhz}) of the appendix A. 

As before, without any loss of generality, we can consider equal 
time correlators where the boundary points are separated only in $x_1$ direction. 
The geodesic equations are:
\begin{equation}
\label{geo1}
 -\frac{2\dot{z}}{z}+\alpha t^{-1+2\alpha}\dot{x}^2=f(t),
\end{equation}
\begin{equation}
\label{geo2}
 \ddot{x}+\frac{2\alpha}{t}\dot{x}-\frac{2}{z}\dot{x}\dot{z}=f(t)\dot{x},
\end{equation}
\begin{equation}
\label{geo3}
\frac{\left(z^6-z_0^6\right) \left(-\dot{x}^2 \left(z^6-z_0^6\right) t^{2 \alpha}+z^6+z z_0^6 \ddot{z}-z_0^6\right)+\dot{z}^2
   \left(z_0^6-4 z^6\right) z_0^6}{z_0^6 \left(z^7-z z_0^6\right)} = f(t)\dot{z},
\end{equation}
where for notational simplicity we have denoted $x_1$ as $x$, avoided the bars from the variables and also called $\alpha_1$ as 
$\alpha$. In the above equations, $f(t)$ is a function of $t$.

Now substituting $f(t)$ from (\ref{geo1}) into (\ref{geo2}) and (\ref{geo3}) we obtain:
\begin{equation}
\label{x-kasner}
 \ddot{x}t=\alpha\dot{x}\left(-2+t^{2\alpha}\dot{x}^2\right),
\end{equation}
\begin{equation}
\label{z-kasner}
\left(z^6-z_0^6\right) \left(z^6+z_0^6 z \ddot{z}-z_0^6\right)-\dot{z}^2 z_0^6 \left(2 z^6+z_0^6\right)
- \dot{x}^2 \left(z^6-z_0^6\right) t^{2 \alpha -1} \left(t \left(z^6-z_0^6\right)+\alpha  z \dot{z} z_0^6\right) = 0.
\end{equation}

We further concentrate on the case $\alpha = \frac{1}{3}$ to see a parallel with the case of the Kasner example we considered in the previous subsection.
For other positive $\alpha$, qualitative behaviour of the correlator remains same.
Equation (\ref{x-kasner}) can be solved analytically. We substitute the solution in (\ref{z-kasner}) and re-express the $z$-equation as a differential equation in $x$.

\begin{eqnarray}
\label{xz-kasner}
&& \left(x^2-9\right)^3 z^{12}+z_0^{12} \left[81 \left\{\left(x^2-9\right) z z''+\left(x^2-9\right) z'^2-2 x z
   z'\right\}+\left(x^2-9\right)^3\right] \nonumber \\
&+& z_0^6 z^6 \left[-81 \left(x^2-9\right) z z''+162 z' \left\{\left(x^2-9\right) z'+x z\right\}-2
   \left(x^2-9\right)^3\right] = 0 \;. 
\end{eqnarray}

Unlike its counterpart in Kasner-AdS, this equation cannot be solved analytically.
However we do find numerical solutions implementing the boundary conditions, namely,
\begin{itemize}
\item $\frac{dz}{dx} = 0$ at the turning point, $t= t^{*}$ of the geodesic.
\item $z=0$ at $t=t_0$.
\end{itemize}
Further, the geodesic length can be written as
\begin{equation}
\label{Lsol}
 L= \frac{2}{9}
   \int_{0}^{x_0-\delta} {\frac{dx}{ z(x)}} \left(x \sqrt{9-x^2} \sqrt{\frac{81 z_0^6 z'(x)^2}{x^2 \left(9-x^2\right)
   \left(z_0^6-z(x)^6\right)}+\frac{9}{x^2}-1} \, \right) \;.
\end{equation}
Here $x_0$ is related to the fixed time-slice $t_0$ at the boundary through the solution of
(\ref{x-kasner}). 
\begin{equation}
\label{x0t0sol}
x_0 = \pm 3 \sqrt{1- t_0^{2\over3}}.
\end{equation}
Coordinates in (\ref{Lsol}) and (\ref{x0t0sol}) are all scaled coordinates as per (\ref{susix}) so that the turning point is now at $t=1$. 
$\delta$ is a sharp cut-off in $x$ and signifies the UV cut-off near AdS boundary. The singularity 
$\delta = 0$ can however be taken care of by subtracting from it the corresponding length in AdS with the same UV cut-off, $\delta$, 
namely

\begin{equation}
L_{AdS} = 2 \, \log \left[\frac{t_0^{\frac{1}{3}}}{z(x_0-\delta)}\right].
\end{equation}
 In figure 1, we plot $e^{-L_{reg}}$ as a function of $t_0$ where 
 \begin{equation}
 L_{reg} = L - L_{AdS}.
 \end{equation}
\begin{figure}[H]
\label{fig}
 \centering
 \includegraphics[width=.75\textwidth]{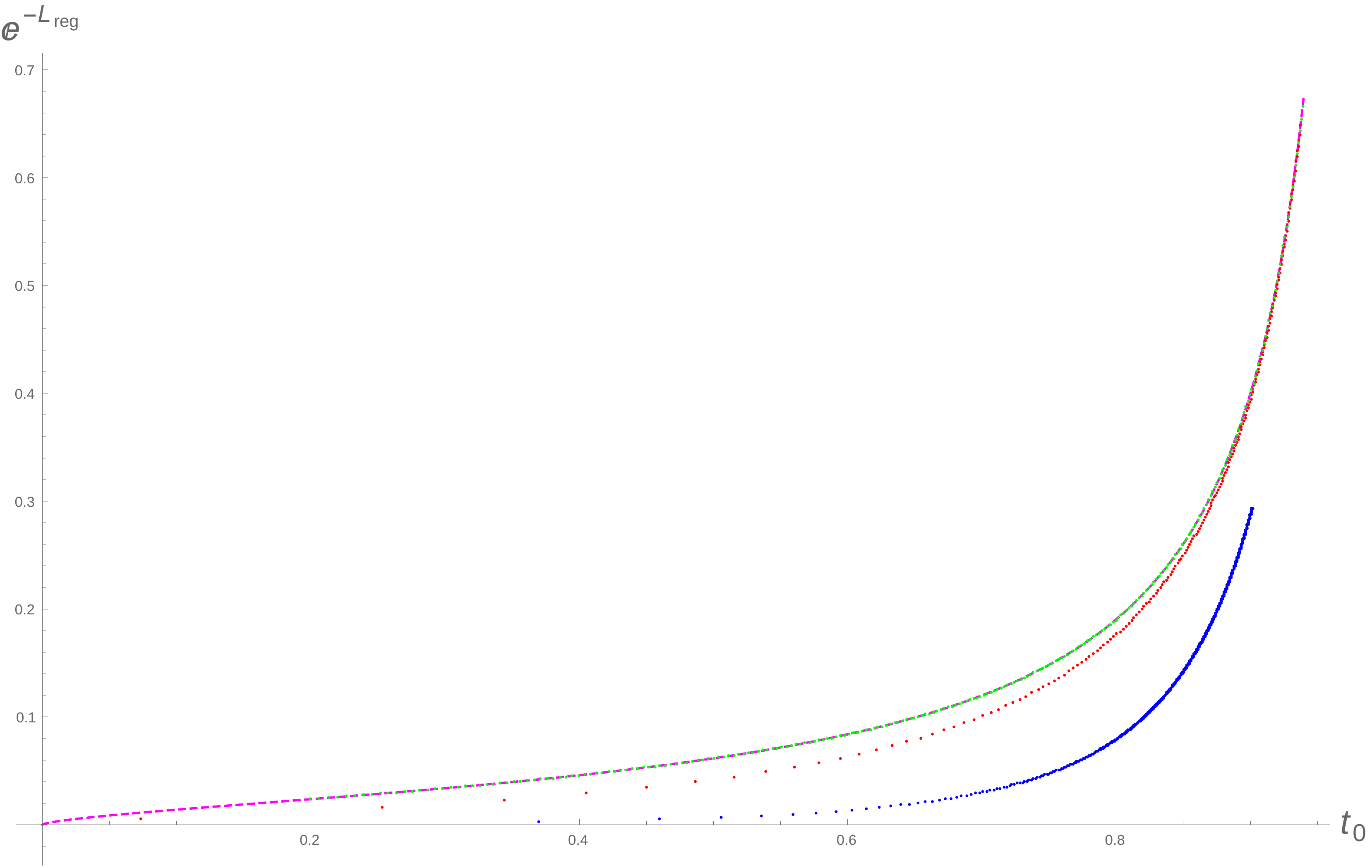}
 \caption{Plot of $e^{-L_{reg}}$ vs $t_0$ for  $AdS_7$ Kasner soliton. Blue, red and green curves are 
for $z_0 = .5, 1$ and $10$ respectively. For comparison, we also showed, in magenta, $e^{-L_{reg}}$ for
pure $AdS_7$ Kasner solution.}
\end{figure}

The results are plotted in figure 2. The correlator qualitatively shows a behaviour similar to the one we argued for the 
Kasner geometry. 
We see here that the function  $e^{-L_{reg}}$ goes to zero smoothly as we tune $t_0 \rightarrow 0$. We further see from 
the figure that, as $z_0$ increases, the plots more and more resemble that of Kasner-AdS. This is expected. As the
point $z_0$ moves away from the boundary, the correlator sense less of the bulk solitonic geometry.

Finally, we note that in the context of AdS/CFT, it is also possible to construct geometry which breaks the underlying scaling 
symmetry. For example, one can consider \cite{Townsend:2003fx}
\begin{equation}
ds^2 = \frac{1}{z^2} \Big[ - dt^2 + dz^2 + \sum_{i=1}^n a_i^2 (t) dx_i^2 + \sum_{j =1}^m b_j^2(t) d\Sigma^2\Big],
\end{equation}
where, $\Sigma$ represents a $m$ dimensional hyperbolic manifold. The metric has cosmological singularities and,
owing to the complicated time-dependence of $a$ and $b$, it breaks the scaling symmetry. It would be interesting
to compute an appropriate boundary correlator for this geometry. We shall report about such 
scaling-violating cosmological solutions and their implications in a later issue \cite{inprep}.

\section*{Acknowledgments}
We thank Sumit Das, Netta Engelhardt, Suman Ganguly, Thomas Hertog, K. Narayan and Amitabh Virmani for  for illuminating discussions and/or 
fruitful communications.

\appendix
\section{Nod from Supergravity}
In this appendix we elaborate on how the time-dependent solutions discussed in previous sections inherit a supergravity origin. The solutions we talk about here 
solves the equations of motion of supergravity theories in $10$ and $11$ dimensions.
\subsection*{Solutions in $D=10$}
We first review the time-dependent solutions from $10$D type IIB supergravity. We primarily concentrate on the Bosonic part of the theory, since for 
generic time-dependent solutions supersymmetry is explicitly broken.
The equations of motion following from the relevant part of standard IIB 
supergravity action\footnote{We impose the self-duality condition of the $5$-form field strength at the level of equation of motion.} 
\begin{equation}
 \label{IIB}
 S_{IIB} = -\frac{1}{16 \pi G_{10}}\int d^{10}x \sqrt{-g}
 \left(R - \frac{1}{2}\partial^\mu \phi \partial_\mu \phi - \frac{1}{2\times5!} F_5^2 \right)\;.
\end{equation}
has the forms:
\begin{eqnarray}
&&R^{\mu}_{\nu} = \frac{1}{2} \partial^\mu \phi \partial_{\nu} \phi + \frac{1}{2 \times 5!} (5 F^{\mu \xi_2...\xi_5}
 F_{\nu \xi_2...\xi_5}- \frac{1}{2} \delta^\mu_\nu F_5^2),\nonumber\\
&&\partial_\mu({\sqrt{g}} F^{\mu \xi_2...\xi_5}) =0, \nonumber\\
&&\nabla^2\phi =0.
\label{2beom}
\end{eqnarray}

\begin{itemize}
\item It was shown in \cite{Banerjee:2013jn} that these equations are solved by the following metric and gauge field configuration:
\begin{eqnarray}
&&ds^2 = \Big(1 + \frac{l^4}{r^4}\Big)^{-\frac{1}{2}} \Big[ - dt^2 + \sum_{i=1}^3t^{2 p_i} dx_i^2\Big] + \Big(1 + \frac{l^4}{r^4}\Big)^{\frac{1}{2}}
\Big[dr^2 + r^2 d\Omega_5^2\Big],\nonumber\\
&&F_{tx_1x_2x_3r} = \frac{2\sqrt{2} l^4t^{p_1+p_2+p_3}r^3}{(l^4 + r^4)^2},
~~~~~
 F_{ijklm} = \sqrt{-g} \,\epsilon_{tx_1x_2x_3rijklm}\, F^{tx_1x_2x_3r}
\nonumber\\
&&\phi =0,
\label{timev}
\end{eqnarray}
provided the exponents, $p_i$, satisfy Kasner conditions, namely
\begin{equation}
\sum_{i=1}^3 p_i = 1~~~{\rm and}~\sum_{i=1}^3 p_i^2 = 1
\label{restric}
\end{equation}
Here, $i, j, k, l, m$ are the indices on $S^5$.

In the near horizon limit, $r \rightarrow 0$, the metric 
reduces to
\begin{equation}
ds^2 = -\frac{r^2}{l^2} dt^2 + \frac{l^2}{r^2} dr^2 + r^2( 
t^{2\alpha} dx^2 + t^{2\beta} dy^2 + t^{2\gamma} dz^2) + l^2 d\Omega_5^2,
\label{kasads}
\end{equation}
with 
\begin{equation}
F_{txyzr} = \frac{4 t r^3}{l^4}, ~~{\rm giving ~potential}~C_{txyz} = \frac{ t r^4}{l^4}.
\end{equation}
We call it a Kasner-$AdS_5$ solution as we discussed under the class of solutions, (\ref{suone}){\footnote {This form is related to the form given in 
(\ref{suone}) by a coordinate transformation, $r=\frac{1}{z}$}}.

\item Kasner solutions sourced by scalar fields can also be realised, likewise, from the same 
supergravity set-up. The scalar field profile, however, in this case gets an interpretation of
stiff matter on the brane configuration in question.

\end{itemize}

\subsection*{Solutions in $D=11$}

In \cite{Banerjee:2013jn} it was also discussed that extremal M5 brane solutions with Kasner-like time dependent scaling 
of transverse spatial coordinates is a solution of (Bosonic sector of) $D=11$ supergravity with time-dependent gauge fields:
\begin{eqnarray}
\label{metric11d}
ds^2 &=& \left(1 + \frac{l^3}{r^3}\right)^{-\frac{1}{3}} \left[-dt^2 + \sum_{i=1}^5 t^{2 p_i} dx_i^2\right] \nonumber \\
&+& \left(1 + \frac{l^3}{r^3}\right)^{\frac{2}{3}} \left[dr^2 + r^2 d\Omega_4^2\right],
\end{eqnarray}
along with
\begin{equation}
\label{gauge11d}
F_{t x_1 x_2 x_3 x_4 x_5 r} = \frac{3 \ l^3 \ t \ r^2}{(l^3 + r^3)^2},
\end{equation}

where the exponents, $p_i$, satisfy Kasner conditions, namely
\begin{equation}
\sum_{i=1}^5 p_i = 1~~~{\rm and}~\sum_{i=1}^5 p_i^2 = 1.
\label{restricM5}
\end{equation}
In the near horizon limit, i.e. $r\rightarrow 0$, the metric and the non-zero component of the 
form field reduce to the forms :
\begin{eqnarray}
\label{nehor} 
ds^2 &=& \frac{r}{l} \left[-dt^2 + t^{2 \alpha_1} dx_1^2 
+ t^{2 \alpha_2} dx_2^2 + t^{2 \alpha_3} dx_3^2 
+ t^{2 \alpha_4} dx_4^2 + t^{2 \alpha_5} dx_5^2\right] \nonumber \\
&+& \frac{l^2}{r^2} \left[dr^2 + r^2 d\Omega_4^2\right],\nonumber  \\
F_{t x_1 x_2 x_3 x_4 x_5 r} &=& \frac{3 \ t \ r^2}{l^3},
\end{eqnarray}
Through a change of coordinate, 
\begin{equation}
\label{w-coord}
w^2 = \frac{r}{l^3}. 
\end{equation}
the metric in (\ref{nehor}) further takes the form :
\begin{equation}
\label{nehor1} 
ds^2 = \frac{w^2}{4 l^2}\left(-d\bar{t}^2 + \bar{t}^{2 \alpha_1} d\bar{x}_1^2 
+ \bar{t}^{2 \alpha_2} d\bar{x}_2^2 + \bar{t}^{2 \alpha_3} d\bar{x}_3^2 
+ \bar{t}^{2 \alpha_4} d\bar{x}_4^2 + \bar{t}^{2 \alpha_5} d\bar{x}_5^2 \right) + 
4 \ l^2 \frac{dw^2}{w^2} + l^2 d\Omega_4^2,
\end{equation}
where $\bar{x}_i$ and $\bar{t}$ are suitably scaled versions of the coordinates, ${x}_i$ and ${t}$ 
respectively. This space we call $KAdS_7 \times S^4$. This solution also belongs to the category
discussed under the general form, (\ref{suone}).

However, there also exists another negative energy solution of the same supergravity sector.
At the level of solutions, such negative energy solutions are obtained through 
a double analytic continuation of the time and the ``p''-th world-volume coordinate of a non-extremal p-brane
solution {\footnote{Note, non-extremal M5 branes with Kasner-like world-volume are not solutions of $11$D 
supergravity}}.
\begin{equation}
t \rightarrow i \theta,  \hspace{0.5in} x^p \rightarrow i t
\end{equation}
These solutions are known as soliton solutions in literature \cite{Horowitz:1998ha}. In the near horizon
limit, one gets the so called AdS solitons which are energetically favoured and hence a more suitable candidate to
study the boundary gauge theory.

$t$ being the time coordinate and $\theta$, a periodic angular coordinate, this double analytic continuation 
amounts to changing the asymptotic 
topology $R^{p}$ of the parent p-brane configuration to $R^{p-1} \times S^1$.
Next we look in detail the case in $11$-D supergravity when the AdS solitons have time-dependent world-volume.

The generic action for  the Bosonic part of $d=11$ supergravity is 
\begin{equation}
 \label{11dsugra}
 S_{11d} = -\frac{1}{2 \ \kappa_{11}^2}\int d^{11}x \sqrt{-g}
 \left(R - \frac{1}{48}  \ F_4^2\right)\;,
\end{equation}

The equations of motion arising from  (\ref{11dsugra}) admits the solitonic solution:
\begin{eqnarray}
\label{solM5}
ds^2 &=& \left(1 + \frac{l^3}{r^3}\right)^{-\frac{1}{3}} \left[-dt^2 + \sum_{i=1}^4 t^{2 p_i} dx_i^2  + 
\gamma\left(r\right) d\theta^2\right] \nonumber \\
&+& \left(1 + \frac{l^3}{r^3}\right)^{\frac{2}{3}} \left[\frac{1}{\gamma\left(r\right)} dr^2 
+ r^2 d\Omega_4^2\right],
\end{eqnarray}
where $\gamma(r) = 1-\frac{r_0^3}{r^3}$,

and the gauge field is given by :
\begin{equation}
\label{gaugesolM5}
F_{\theta x_1 x_2 x_3 x_4 t r} = \frac{3 \sqrt{l^3 + r_0^3} \ l^{\frac{3}{2}} \ t \ r^2}{(l^3 + r^3)^2}.
\end{equation}
iff the exponents, $p_i$'s satisfy Kasner condition, namely
\begin{equation}
\sum_{i=1}^4 p_i = 1~~~{\rm and}~\sum_{i=1}^4 p_i^2 = 1
\label{restricM5soliton}
\end{equation}.

We call this solution a M5-soliton.

In near horizon limit, the M5-Kasner soliton solution takes the form
\begin{eqnarray}
\label{nehorsol} 
ds^2 &=& \frac{r}{l} \left[-dt^2 + \sum_{i=1}^4 t^{2 p_i} dx_i^2 + \gamma(r) d\theta^2\right]
+ \frac{l^2}{r^2} \left[\frac{1}{\gamma(r)}dr^2 + r^2 d\Omega_4^2\right],\nonumber  \\
&& F_{t x_1 x_2 x_3 x_4 x_5 r} = \frac{3 \ t \ r^2}{l^3},
\end{eqnarray}
We work in the same coordinates defined in (\ref{w-coord}).
In these coordinates the metric takes the form
\begin{equation}
\label{M5nh}
 ds^2 = \frac{w^2}{4 l^2}\left[-d\bar{t}^2 + \sum_{i=1}^4 {\bar t}^{2 p_i} d{\bar x}_i^2 +\left(1-\frac{w_0^6}{w^6}\right)
d\bar{\theta}^2\right] 
+ 4 \ l^2 \left(1-\frac{w_0^6}{w^6}\right)^{-1} \frac{dw^2}{w^2} + l^2 d\Omega_4^2,
\end{equation}
where 
$w_0=\frac{r_0}{l^3}$. $\bar{x}_i$ and $\bar{t}$ are suitably scaled versions of the coordinates, ${x}_i$ and ${t}$ respectively. 
Note here, additionally, $\theta$ is also rescaled to $\bar{\theta}$ and hence the period of 
$\bar{\theta}$ has to be adjusted accordingly. 

Employing the coordinate transformation $\omega = \frac{4 l^2}{z}$,
the $AdS_7$ part of the metric reduces to the familiar form :
\begin{eqnarray}
\label{M5nhz}
 ds^2 &=& \frac{1}{z^2}\left[-d\bar{t}^2 + \bar{t}^{2 \alpha_1} d\bar{x}_1^2 
+ \bar{t}^{2 \alpha_2} d\bar{x}_2^2 + \bar{t}^{2 \alpha_3} d\bar{x}_3^2 
+ \bar{t}^{2 \alpha_4} d\bar{x}_4^2 +\left(1-\frac{z^6}{z_0^6}\right)
d\bar{\theta}^2 + \left(1-\frac{z^6}{z_0^6}\right)^{-1} dz^2\right]\nonumber\\
\end{eqnarray}
This is the same metric as in (\ref{kasol}).

\section{Solving of the geodesic equations}

Here we discuss a way to solve equations (\ref{suthree}). It is best to define a new time coordinate
$\eta$ such that
\begin{equation}
\eta = \int \frac{dt}{a^2(t)}.
\label{appone}
\end{equation}
The first equation in (\ref{suthree}) then reduces to
\begin{equation}
\frac{d^2x}{d\eta^2} - \frac{1}{a^3} \frac{da}{d\eta} \Big(\frac{dx}{d\eta}\Big)^3 = 0.
\end{equation}
Integrating twice, we have
\begin{equation}
x(\eta) = \pm \int \frac{a(\eta)d\eta}{\sqrt{c_1 a(\eta)^2 + 1}} + c_2.
\label{appthree}
\end{equation}
Here $c_1$ and $c_2$ are the integration constants. Now $c_1$ can be fixed using the boundary condition:
at the turning point $\eta = \eta^*$, $dx/dt$ or equivalently $dx/d\eta$ is infinity. This gives 
\begin{equation}
c_1 = - \frac{1}{a^2(\eta^*)}.
\end{equation}
Substituting this in (\ref{appthree}), we can easily integrate the expression. The result is
\begin{eqnarray}
x(\eta) = \frac{(1 - 2 p)^{\frac{1-p}{1 - 2p}} \eta^{\frac{1-p}{1 - 2p}}   }{1 - p}
{}_2F_1\Big(\frac{1}{2}, \frac{1}{2}(-1 + \frac{1}{p}), \frac{1+p}{2p}, \eta^{\frac{2p}{1-2p}} \eta^{* 
-\frac{2p}{1-2p}} \Big)
+c_2,
\label{appfive}
\end{eqnarray}
where we have used the fact that $a(t) = t^p$. Further, $c_2$ can be fixed using 
$x(\eta) = 0$ at $\eta = \eta^*$. This gives
\begin{equation}
c_2 = - \frac{{\sqrt{\pi}}(1 - 2p)^{\frac{p}{1-2p}} \eta^{* \frac{1-p}{(1-2p)}} \Gamma(-\frac{1}{2} + 
\frac{1}{2p})}
{\Gamma(-1 + \frac{1}{2p})}.
\end{equation}
Going over to the $t$ variable, we can write the above equation as
\begin{equation}
x(t) = \frac{t^{1-p}}{1-p} {}_2F_1\Big(\frac{1}{2}, \frac{1}{2}(-1 + \frac{1}{p}), \frac{1+p}{2p}, 
(\frac{t}{t^*})^{2p} \Big)  - \frac{{\sqrt \pi} \Gamma(-\frac{1}{2} +
\frac{1}{2p}) t^{* (1-p)}}{ (1- 2p){\Gamma(-1 + \frac{1}{2p})}}.
\label{appa}
\end{equation}

Now we turn to the second equation of (\ref{suthree}). Defining $K = z z^\prime$, we first rewrite it as
\begin{equation}
\frac{dK}{dt} - a a^\prime {x^\prime}^2 K + a^2 {x^\prime}^2 -1 =0.
\end{equation}
After going to $\eta$ variable, above equation can easily be integrated. This gives
\begin{equation}
K = \frac{c_1 a}{\sqrt{c_1 a^2 +1}}\int \frac{a^3}{\sqrt{c_1 a^2 +1}} d\eta + c_3,
\end{equation}
where $c_3$ is an integration constant and $c_1$ has been defined earlier. More explicitly, we 
get
\begin{equation}
K = - \frac{ (1 - 2 p)^{\frac{2p}{1 - 2p}} \eta^{\frac{p}{1 - 2p}} }{\sqrt{\eta^{* \frac{2p}{1-2p}} - 
\eta^{\frac{2p}{1-2p}}}} \Big[\int d\eta \frac{ \eta^{\frac{3p}{1-2p}}}{ \sqrt{\eta^{* \frac{2p}{1-2p}} -
\eta^{\frac{2p}{1-2p}}}}+ c_3\Big].
\end{equation}
It is easy to show that, for $dz/dx$ to vanish at $\eta = \eta^*$, the expression inside the brackets 
has to vanish. This, in turn, fixes $c_3$. This gives 
\begin{eqnarray}
K = &&- \frac{ (1 - 2 p)^{\frac{2p}{1 - 2p}} \eta^{\frac{p}{1 - 2p}} }{\sqrt{\eta^{* \frac{2p}{1-2p}} -
\eta^{\frac{2p}{1-2p}}}}\Big[ 
(1 - 2 p) \eta^{\frac{1-p}{1-2p}} \{ \eta^{*\frac{p}{1-2p}} {}_2F_1\Big(\frac{1}{2}, \frac{1}{2}(-1 + \frac{1}{p}), 
\frac{1+p}{2p},
(\frac{\eta}{\eta^*})^{\frac{2p}{1-2p}}\Big) \nonumber\\
&&- {\sqrt{ \eta^{* \frac{2p}{1-2p}} - \eta^{ \frac{2p}{1-2p}} }}\}
- \frac{2 p {\sqrt {\pi}} \eta^{* \frac{1}{1-2p}} \Gamma(\frac{1+p}{2p}) } {\Gamma(\frac{1-2p}{2p}) } \Big].
\end{eqnarray}
Now using the property
\begin{equation}
{}_2F_1 (a, b, c, z) = (1 - z)^{c - a - b} {}_2F_1(c -a, c - b, c, z),
\end{equation}
the expression for $K$ can be simplified to
\begin{eqnarray}
K = &&(1 - 2 p)^{\frac{1}{1-2p}} \eta^{\frac{1}{1-2p}}\Big[1 -
{}_2F_1\Big(\frac{1}{2p}, 1, \frac{1+p}{2p}, (\frac{\eta}{\eta^*})^{\frac{2p}{1-2p}}\Big)\Big] \nonumber\\
&&+ \frac{   2 p (1 -2p)^{\frac{2p}{1-2p}} \eta^{\frac{p}{1-2p}} {\sqrt \pi} \eta^{* \frac{1}{1- 2p}} 
\Gamma(\frac{1+p}{2p}) }{ \Gamma(\frac{1-2p}{2p}) 
{\sqrt{ \eta^{* \frac{2p}{1-2p}} -\eta^{\frac{2p}{1-2p}} }} }.
\end{eqnarray}
This gives
\begin{eqnarray}
z^2 = && \int 2 (1 - 2 p)^{\frac{1 + 2p}{1-2p}} \eta^{\frac{1+ 2p}{1-2p}}\Big[1 -
{}_2F_1\Big(\frac{1}{2p}, 1, \frac{1+p}{2p}, (\frac{\eta}{\eta^*})^{\frac{2p}{1-2p}}\Big)\Big] d\eta\nonumber\\
&& + \frac{    4 {\sqrt \pi} p (1 - 2 p)^{ \frac{4p}{1 - 2p}} \eta^{* \frac{1}{1-2p}}
\Gamma (\frac{1+p}{2p})  
}
{  \Gamma(\frac{1-2p}{2p}) }  \int \frac{\eta^{ \frac{3 p}{1-2p}}} {
{\sqrt{ \eta^{* \frac{2p}{1-2p}} -\eta^{\frac{2p}{1-2p}} }} } d\eta + \tilde c
\end{eqnarray}
where $\tilde c$ is a constant. Carrying out the integrations, we finally get
\begin{eqnarray}
z =&& \Big[(1 - 2 p)^{\frac{2}{1-2p}} \eta^{\frac{2}{1-2p}} \Big[
1 - {}_3F_2\Big( \{1, \frac{1}{2p}, \frac{1}{p}\}, \{\frac{1}{2} + \frac{1}{2p}, 1 + \frac{1}{p}\}, 
(\frac{\eta}{\eta^*})^{\frac{2p}{1-2p}}\Big)\Big]\nonumber\\
&&+ \frac{4{\sqrt \pi} p (1 - 2p)^{\frac{1+2p}{1-2p}} \eta^{* \frac{1}{1-2p}} \Gamma(\frac{1+p}{2p}) 
\eta^{\frac{1-p}{1-2p}} } {\Gamma(\frac{1-2p}{2p})}
\Big[ - {\sqrt{ \eta^{* \frac{2p}{1-2p}} -\eta^{\frac{2p}{1-2p}} }}\nonumber\\
&&+ \eta^{* \frac{p}{1-2p}}
{}_2F_1\Big(\frac{1}{2}, \frac{1-p}{2p}, \frac{1+p}{2p}, (\frac{\eta}{\eta^*})^{\frac{2p}{1-2p}}\Big)\Big]
+ \tilde c\Big]^{\frac{1}{2}}.
\end{eqnarray}
In terms of variable $t$, we therefore find
\begin{eqnarray}
z(t) =&& \Big[t^2\Big[ 1 - {}_3F_2\Big( \{1, \frac{1}{2p}, \frac{1}{p}\}, \{\frac{1}{2} + \frac{1}{2p}, 1 + 
\frac{1}{p}\}, 
(\frac{t}{t^*})^{2p}\Big)\Big] \nonumber\\ 
&&+\frac{4{\sqrt \pi} p \Gamma(\frac{1+p}{2p})
t^{*1+ p} t^{1-p} } { (1-2p)\Gamma(\frac{1-2p}{2p})}
\Big[ {}_2F_1\Big(\frac{1}{2}, \frac{1-p}{2p}, \frac{1+p}{2p}, (\frac{t}{t^*})^{2p}\Big)
- {\sqrt {1 - (\frac{t}{t^*})^{2p}}}\Big] 
+ \tilde c\Big]^{\frac{1}{2}}.
\label{appb}
\end{eqnarray}
Equations (\ref{appa}) and (\ref{appb}) are used in the main text.

\bibliographystyle{JHEP}

\end{document}